\documentstyle[preprint,version2,aps,epsf,prl]{revtex}

\def\3he{{$^3${\rm He}}}

\def\slD{\raise.15ex\hbox{$/$}\kern-.57em\hbox{$D$}}
\def\dsl{\raise.15ex\hbox{$/$}\kern-.57em\hbox{$\Delta$}}
\def\slp{{\raise.15ex\hbox{$/$}\kern-.57em\hbox{$\partial$}}}
\def\nsl{\raise.15ex\hbox{$/$}\kern-.57em\hbox{$\nabla$}}
\def\sla{\raise.15ex\hbox{$/$}\kern-.57em\hbox{$\rightarrow$}}
\def\slla{\raise.15ex\hbox{$/$}\kern-.57em\hbox{$\lambda$}}
\def\gtwid{\raise.3ex\hbox{$>$\kern-.75em\lower1ex\hbox{$\sim$}}}
\def\ltwid{\raise.3ex\hbox{$<$\kern-.75em\lower1ex\hbox{$\sim$}}}

\def\12{{1\over2}}

\def\part{\partial}

\def\bethlogo{\vbox{\bf \line{\hrulefill} 
    \kern-.5\baselineskip 
    \line{\hrulefill\phantom{ ELIZABETH A. MASON }\hrulefill} 
    \kern-.5\baselineskip 
    \line{\hrulefill\hbox{ ELIZABETH A. MASON }\hrulefill} 
    \kern-.5\baselineskip 
    \line{\hrulefill\phantom{ 1411 Chino Street }\hrulefill} 
    \kern-.5\baselineskip 
    \line{\hrulefill\hbox{ 1411 Chino Street }\hrulefill} 
    \kern-.5\baselineskip 
    \line{\hrulefill\phantom{ Santa Barbara, CA 93101 }\hrulefill} 
    \kern-.5\baselineskip 
    \line{\hrulefill\hbox{ Santa Barbara, CA 93101 }\hrulefill}
    \kern-.5\baselineskip 
    \line{\hrulefill\phantom{ (805) 962-2739 }\hrulefill} 
    \kern-.5\baselineskip 
    \line{\hrulefill\hbox{ (805) 962-2739 }\hrulefill}}}
\def\lisalogo{\vbox{\bf \line{\hrulefill} 
    \kern-.5\baselineskip 
    \line{\hrulefill\phantom{ LISA R. GOODFRIEND }\hrulefill} 
    \kern-.5\baselineskip 
    \line{\hrulefill\hbox{ LISA R. GOODFRIEND }\hrulefill} 
    \kern-.5\baselineskip 
    \line{\hrulefill\phantom{ 6646 Pasado }\hrulefill} 
    \kern-.5\baselineskip 
    \line{\hrulefill\hbox{ 6646 Pasado }\hrulefill} 
    \kern-.5\baselineskip 
    \line{\hrulefill\phantom{ Santa Barbara, CA 93108 }\hrulefill} 
    \kern-.5\baselineskip 
    \line{\hrulefill\hbox{ Santa Barbara, CA 93108 }\hrulefill}
    \kern-.5\baselineskip 
    \line{\hrulefill\phantom{ (805) 962-2739 }\hrulefill} 
    \kern-.5\baselineskip 
    \line{\hrulefill\hbox{ (805) 962-2739 }\hrulefill}}}

\def\low#1{\lower.5ex\hbox{${}_#1$}}
\def\ltwid{\raise.3ex\hbox{$<$\kern-.75em\lower1ex\hbox{$\sim$}}}

\def\psl{\raise.15ex\hbox{$/$}\kern-.57em\hbox{$\partial$}}
\def\partt{\raise.15ex\hbox{$\widetilde$}{\kern-.37em\hbox{$\partial$}}}
\def\parts{\raise.15ex\hbox{$/$}{\kern-.6em\hbox{$\partial$}}}
\def\nablas{\raise.15ex\hbox{$/$}{\kern-.6em\hbox{$\nabla$}}}
\def\oprod{\hbox{$\rm O$}{\kern -0.8em\hbox{$\Pi$}}}
\def\partw#1{\raise.15ex\hbox{$/$}{\kern-.6em\hbox{${#1}$}}}

\def\gtappr{{{\lower4pt\hbox{$>$} } \atop \widetilde{ \ \ \ }}}
\def\ltappr{{{\lower4pt\hbox{$<$} } \atop \widetilde{ \ \ \ }}}

\def\topppageno1{\global\footline={\hfil}\global\headline
={\ifnum\pageno<\firstpageno{\hfil}\else{\hss\twelverm --\ \folio
\ --\hss}\fi}}

\def\toppageno2{\global\footline={\hfil}\global\headline
={\ifnum\pageno<\firstpageno{\hfil}\else{\rightline{\hfill\hfill
\twelverm \ \folio
\ \hss}}\fi}}

\def\ltdash{\raise-1.8pt\hbox{$\scriptscriptstyle |$}}

\def\1{{\bf 1}}
\def\2{{\bf 2}}

\def\ell{{\it l } {\rm n}}

\def\cx2{\sqrt{c^2_x+c^2_y}}

\def\gkk{\gamma _{\vec k}}
\def\gk2{\gkk ^2}

\def\gtappr{{{\lower4pt\hbox{$>$} } \atop \widetilde{ \ \ \ }}}
\def\ltappr{{{\lower4pt\hbox{$<$} } \atop \widetilde{ \ \ \ }}}

\def\dsp{\displaystyle}
\def\pbar{{\partial\kern-1.2ex\raise0.25ex\hbox{/}}}

\def\dsp{\displaystyle}

\def\1{{\bf 1}}
\def\2{{\bf 2}}

\def\ell{{\it l } {\rm n}}

\def\cx2{\sqrt{c^2_x+c^2_y}}

\def\gkk{\gamma _{\vec k}}
\def\gk2{\gkk ^2}
\def\gtappr{{{\lower4pt\hbox{$>$} } \atop \widetilde{ \ \ \ }}}
\def\ltappr{{{\lower4pt\hbox{$<$} } \atop \widetilde{ \ \ \ }}}

\def\thickra{\hbox{\raise0.2pt\hbox{{$\bf >\mkern-13mu>\mkern-13mu>$}}}}
\def\thickrarrow{\hbox{\raise0.28pt\hbox{{$\bf >\mkern-13mu>\mkern-13mu>$}}}}

\def\sl{\it}

\catcode`\@=11       
\def\listzigurename{List of Zigures}

\def\zigurename{Zigure}

\def\listofzigures{\section*{\listzigurename
    \@mkboth{\uppercase{\listzigurename}}{\uppercase{\listzigurename}}}%
  \@starttoc{lof}}

\def\l@zigure{\@dottedtocline{1}{1.5em}{2.3em}}

\let\l@table\l@zigure

\newcounter{zigure}
\def\thezigure{\@arabic\c@zigure}

\def\fps@zigure{tbp}
\def\ftype@zigure{1}
\def\ext@zigure{lof}
\def\fnum@zigure{\zigurename~\thezigure}
\def\zigure{\@float{zigure}}
\let\endzigure\end@float
\@namedef{zigure*}{\@dblfloat{zigure}}
\@namedef{endzigure*}{\end@dblfloat}
\catcode`\@=12
\begin{document}
\onecolumn
\vskip 0.2truein
\begin{title}
Nonlinear Susceptibility: A Direct Test of the Quadrupolar 
Kondo Effect in
\centerline{$UBe_{13}$}
\end{title}

\author{ A.P. Ramirez,$^1$ P. Chandra,$^2$ P. Coleman,$^{3}$ Z.Fisk,$^4$ J.L. Smith,$^4$ and H.R. Ott$^5$}

\begin{instit}$^1$ A.T. \& T. Bell Laboratories
600 Mountain Avenue,
Murray Hill, NJ 07974.

$^2$ NEC Research Institute,
4 Independence Way,  
Princeton, NJ 08540.

$^3$ Serin Physics Laboratory,
Rutgers University,
P.O. Box 849,
Piscataway, NJ 08854.

$^4$ Los Alamos National Laboratory,
            Los Alamos, New Mexico 87545. 

$^5$ Eidg. Tech. Hochschule,
            ETH-H\"onggerberg,
            CH-8093 Z\"urich, Switzerland
\end{instit}

\begin{abstract}
We present the nonlinear susceptibility 
as a direct test of the quadrupolar Kondo 
scenario for heavy fermion behavior, and apply it
to the case of cubic crystal-field symmetry.
Within a single-ion model we compute the nonlinear susceptibility
resulting from low-lying $\Gamma_3$ ($5f^2$) and Kramers ($5f^3$)
doublets.  We find that nonlinear susceptibility measurements 
on single-crystal $UBe_{13}$ are {\sl inconsistent} with a quadrupolar
$(5f^2)$ ground-state of the uranium ion; the experimental data indicate
that the low-lying magnetic excitations of $UBe_{13}$ are predominantly
{\sl dipolar} in character.
\end{abstract}


\newpage

There exist several metallic systems whose novel thermodynamic,
magnetic and transport 
properties
are not adequately described by conventional Fermi liquid theory;
specific examples include the quasi-one dimensional conductors,\cite{oned}
certain actinide heavy fermion materials\cite{1,2,3,4,5} and the layered cuprate
superconductors.\cite{6,7}  The search and characterization of
non-Fermi liquid (NFL) fixed points is thus a topic of active 
research.\cite{3,4,5,6,7}
In this Letter we present an unambiguous experimental test of the
quadrupolar Kondo effect,  a model proposed by
Cox\cite{4} to characterize the NFL behavior observed in the 
cubic three-dimensional heavy fermion material $UBe_{13}$.
We use the nonlinear susceptibility ($\chi_3$) as a direct
probe of low-lying quadrupolar fluctuations, and compute its
behavior within a single-ion model for the case of
cubic crystal-field symmetry; these predictions are then compared
to $\chi_3$ measurements on single-crystal $UBe_{13}$.

Most heavy fermion metals display a dramatic reduction
in resistivity at low temperatures associated with the development
of coherent quasiparticle propagation.  $UBe_{13}$ is atypical, 
undergoing a superconducting
transition directly from a normal state with a large incoherent resitivity\cite{ott0}
of order $140\mu \Omega \ cm$.
The low-temperature dependences of the magnetic
susceptibility\cite{ott0} and the specific heat\cite{ott0} are logarithmic in the approach
to the superconducting transition.
Resistance,\cite{100} specific heat,\cite{101}, susceptibility\cite{102} and
magnetoresistance\cite{103} measurements
indicate that Fermi liquid behavior is restored at low temperatures
under an applied
pressure.  Thus $UBe_{13}$ is a metal with
a {\sl tunable} Fermi temperature $(T_F^*)$ such that $T_F^* < T_c$
at ambient pressure.  The microscopic physics underlying this
suppressed, pressure-dependent\cite{100,101,102,103} and field-dependent\cite{batlogg} $T_F^*$ is a crucial issue for
the characterization of the complex many-body ground-state
of $UBe_{13}$.

Cox\cite{4} has proposed that novel single-ion physics is responsible
for the observed NFL behavior in $UBe_{13}$. The observation of
a well-defined Schottky anomaly\cite{steglich} at $T \sim 180 K$ indicates
that the uranium ion is in a local-moment rather than an intermediate
valence regime;\cite{note} however, it can assume
either the $U^{4+}$ ($5f^2$) or the
$U^{3+}$ ($5f^3$) nominal valence state,\cite{4,steglich}
and neither
quasi-elastic neutron
scattering\cite{neutrons} nor photoemission\cite{ott2,allen} measurements can
unambiguously resolve the crystal-field assignments.  In a cubic
enviroment Cox has identified a non-magnetic {\sl quadrupolar}
($\Gamma_3$) ground-state\cite{4,llw} of the $U^{4+}$ ion.  
He suggests that fluctuations within
this non-Kramers doublet are overscreened by
the conduction electrons; this  quadrupolar Kondo effect then
leads naturally to
a non-Fermi liquid ground-state.\cite{nb} In an alternate scenario,
supported by NMR measurements\cite{doug} consistent with  a $U^{4+}$
valence state,
the low-lying spin excitations are dipolar; the NFL
behavior is attributed to the system's proximity to a $T=0$ quantum
phase transition,\cite{3,5} analogous to that recently observed in
$MnSi$ by Lonzarich and coworkers.\cite{gil}

The nonlinear susceptibility ($\chi_3$) is an ideal test of
the quadrupolar Kondo effect in $UBe_{13}$; it can
distinguish unambiguously between
a low-lying quadrupolar and Kramers crystal-field doublet.
In the paramagnetic state, $\chi_3$ measures 
the leading nonlinearity in the magnetization 
\begin{equation}
M = \chi_1 B + {1 \over 3!} \chi_3 
 B^3+ \dots \label{1} 
\end{equation}
in the direction of the applied field ($B$); it was originally proposed as a 
direct
probe of order-parameter fluctuations in spin glasses.\cite{chalupa}
Morin and Schmitt\cite{ms} extended this technique to non-random spin 
systems,
where they used the nonlinear susceptibility to study quadrupolar 
interactions
in rare-earth intermetallic compounds. 

The most general form for $\chi_3$ in a cubic enviroment is\cite{jeffreys}
\begin{equation}
\chi_3 = \chi_3^{111} + \Delta\chi_3 \Phi(\hat b)
\label{gen}
\end{equation}
where $\Phi(\hat b)$ is the cubic harmonic
\begin{equation}
\Phi(\hat b) = {1\over 2}\left [ 3 (b_x^4 +b_y^4+b_z^4) - 1 \right]
\label{phi}
\end{equation}
and the $b_i\quad (i=1,2,3)$ are the direction cosines of the field.
The numerical factors in $\Phi(\hat b)$ are chosen
so that 
$\Delta \chi_3 \equiv \chi_3^{100} - \chi_3^{111}$; the 
``powder-averaged'' component of the nonlinear susceptibility
is $\overline{\chi_3} = \chi_3^{111} + {7\over 20} \Delta \chi_3$.

The ratio of the two contributions to $\chi_3$ in 
(2) is qualitatively different
for a quadrupolar and a magnetic ground-state.
An isolated  Kramers doublet\cite{ms} results in 
a nonlinear susceptibility 
$\chi_3 = -{{\mu_0^4}\over {3T^3}}$
that is {\sl isotropic}, 
reflecting the negative curvature of the Brillouin 
function.
For an isolated doublet with a {\sl quadrupolar} moment $Q$, the field-dependent
part of the Hamiltonian is $\hat H={1 \over 2} B^2\hat Q_{ab}b_a b_b$, where
$\hat Q_{ab} \propto [ \hat J_a \hat J_b - { 1 \over 3} \delta_{ab}
J(J+1)]$
is the quadrupole operator;\cite{llw} more explicitly
\begin{equation}
\hat H=
{Q B^2\over 2}
\left[
\matrix{ q_{zz} &  q_{xx}- i q_{yy} \cr
  q_{xx}+ i q_{yy} & -q_{zz} \cr
}
\right]\label{equad2x}
\end{equation}
where $q_{aa} = b_a^2 - {1 \over 3}$ ($a=x,y,z$). Diagonalizing
$H$, we find that the splitting of the quadrupolar doublet is given by
\begin{equation}
E^q_{\Gamma \pm} = E_{\Gamma} \pm {QB^2} \sqrt{\Phi (\hat b)\over 4 !}
\label{equad2}
\end{equation}
which yields an {\sl anisotropic} nonlinear susceptibility 
$
\chi_3(\hat b) = {Q^2\over 2T} \Phi(\hat b)
$.
In Cox's model\cite{4} for $UBe_{13}$, there is partial quenching
of $Q$
by the
conduction sea and 
\begin{equation}
\Delta \chi_3 = {{Q^2 \over 2T} f(T/T_0)} = \quad \left\{ 
\begin{array}{rl}
&\dsp{Q\over 2T} \qquad\qquad\quad\quad  T >> T_0\\ 
&\dsp{\alpha Q\over 2T_0} \ln (T_0/T) \qquad T << T_0
\end{array}\right.
\label{chi3}
\end{equation}
where $T_0$ is the ``Bethe ansatz'' Kondo temperature;\cite{21} the exact solution of the two-channel
Kondo model \cite{21,20,22} yields an asymptotic form
for $f(x)$ with an associated value\cite{21} $\alpha = 1/{\pi^2}\approx 0.10$.
Thus the quadrupolar
Kondo hypothesis predicts a $\Delta\chi_3(\hat b)$ that {\sl increases} logarithmically with decreasing temperature.

In this idealized discussion we have neglected the Van Vleck 
contributions to $\chi_3$. In practice, a uranium atom in a
magnetic configuration ($U^{3+}$) with a moment
$
\mu(H) = \mu_0 + { B^2 A(\hat b)\over 3!}
$
will exhibit a small $\Delta \chi_3 = {4 \mu_0 A (\hat b) \over T}$ due to
the  nonlinearity in
$\mu(H)$.
Conversely,
$\chi_3$ for a $U$ ion with a low-lying quadrupolar doublet ($U ^{4+}$) {\sl will}
have an {\sl isotropic} Van Vleck component ($\chi_3^{VV}$)
that has a weak temperature dependence. Despite these additional contributions,
we expect the
anisotropic component of the nonlinear susceptibility
\begin{equation}
{\Delta \chi_3(\hat b) \over \chi_3}   
\equiv {(\chi_3(\hat b)-\chi_3^{111}) \over \chi_3^{111}}
\end{equation}
to be {\sl small} and nearly temperature-independent for a uranium atom with a dipolar ground-state; 
by contrast, ${\Delta \chi_3(\hat b)\over \chi_3}$
should be {\sl large}
and strongly temperature-dependent if the low-lying fluctuations are 
quadrupolar in nature.

Single-ion crystal-field calculations allow us to quantify the preceeding
discussion, and they have been performed for $J=4$ and $J = {9\over 2}$ manifolds
of $f$ orbits in a cubic enviroment.\cite{llw}
The overall energy scale ($W$) and the
level ordering ($x$) have been adjusted to fit the total entropy in
the observed Schottky anomaly,\cite{steglich}
and the resulting energy schemes are displayed in Figure 1.  
The associated $\chi_3(\hat b)$ and $\Delta\chi_3(\hat b)\over \chi_3$
are shown in Figure 2 and 3 respectively,
where the moment has been normalized by a fit to the
measured high-temperature susceptibility;\cite{ott}
the numerical solution of the two-channel Kondo model\cite{21} has been used
to determine the effects of screening in Figure 3.

In order to test the quadrupolar scenario in $UBe_{13}$, we measured $\chi_3$ along the three principal
crystal axes of an oriented single crystal grown from $Al$ flux.
The superconducting transition temperature, a rough measure of the sample quality, 
was found by specific heat to be 
$T_c = 0.75 K$ for this crystal.
Measurements were also performed on a 
a polycrystalline sample with $T_c = 0.96$. Finally a third sample,
an unoriented single crystal with $T_c = 0.48 K$, was studied.
For the $\chi_3$ measurements on the oriented crystal, the orientation was achieved
with a precision of
$\pm 3$ degrees; the data were taken as $M$ vs. $B$
at fixed temperatures up to 4 Tesla in a Quantum Design SQUID magnetometer.  The deviation from linearity was only 
$\sim 2 \%$
at the lowest temperature and the highest field;
it was attributed to the leading nonlinear contribution of $M$ to $\chi_3$.
The magnetization data were fit to the expression 
$M = \chi_0 + \chi_1 B + {1\over 3!}\chi_3 B^3$,
where $\chi_0$ was
included to avoid systematic errors associated with both trapped flux 
in the superconducting solenoid and a small ($\sim 10$ ppm), ubiquitous ferromagnetic
signal which saturated at $\sim 1$ Tesla.
The temperature dependence of
$\chi_3$ is displayed in Figure  4.  The data were typically fit over the region $2 < B < 4$ Tesla;
in this field range, $M/B - \chi_0$ was always linear with
respect to $B^2$.  The linear part, $\chi_1$ (not shown), agrees well with published
values.\cite{ott0} Figure 4 shows the nonlinear susceptibility measured in 
the 111, the 110, and the 100 directions.
We note that the observed $\chi_3$ is both
{\sl negative} and monotically {\sl decreasing} with decreasing temperature; its magnitude
is significantly greater than that predicted
for the quadrupolar scenario (Figure 2a),
but comparable in size at $T \sim 10 K$ with that expected from a dense concentration of
only partially quenched $U$ magnetic doublets.

The observed magnitude and temperature dependence of $\chi_3$ was
similar for the other two samples studied. The measurements on the
polycrystalline sample (Figure 4) provide a crucial
control on our results; here we expect the impurity level
to be low given the relatively high observed
value of $T_c$.
The polycrystalline sample displays
behavior in $\chi_3(T)$ similar to that of the orientation-averaged
single-crystal. This result, combined with the large magnitude of
$\chi_3$, exclude the possibility that the observed $\chi_3$ is due
a residual background of magnetic impurities.

The measured anisotropy in the nonlinear susceptibility (Figure 4 inset) is
small (${\Delta \chi_3 (\hat b) \over \chi_3} \sim 3 \times 10^{-1}$) with a
very weak temperature dependence; moreover it appears,
at the level of one standard deviation,  to have the {\sl opposite}
sign to that expected for the quadrupolar scenario (see Figure 3). 
These results strongly favor a
magnetic model for the uranium ions in $UBe_{13}$ with a low Kondo temperature.
One can try to reconcile these results with the quadrupolar
scenario by invoking a large Van Vleck contribution ($\chi_3^{VV}$); 
it would result from virtual spin or valence fluctuations
into higher lying multiplets of the $U$ ion.
Such a term would scale approximately with $1 \over
\Delta_x$, where $\Delta_x$ is the gap to the higher multiplets.
In order for $\chi_3^{VV} \sim {1\over \Delta_x}$ to be much larger than
$\Delta \chi_3 \sim {1\over T_0}$  we need
$\Delta_x < T_0$, a condition {\sl inconsistent} with
the initial assumption of a well-defined quadrupolar ground-state.

We now return to the possible origins of
NFL behavior in $UBe_{13}$.
Though a single-ion mechanism cannot be ruled
out,\cite{impurity}a canonical Kondo model for the magnetic $U$ ion 
results in a Fermi liquid ground-state. 
Furthermore one expects a system with a low-lying Kramers doublet
to display a reduction in $\gamma \equiv {c_v \over T}$
when $g\mu_B B \sim T_F^*$,
in contrast to that observed\cite{mayer} for $UBe_{13}$.
Thus we conclude that these results {\sl cannot} be explained
within a singe-ion picture picture and require a more sophisticated
approach, possibly one that has an intrinsic pressure- and field-dependent
$T_F^*$.
We are tempted to identify the observed NFL
behavior as a lattice phenomenon, 
possibly attributed to the system's proximity to a $T=0$
quantum phase transition.\cite{3,5} 
Two different types of experiments would clarify this situation.
First, thermodynamic and transport studies on $U_{x}Th_{1-x}Be_{13}$ 
would probe the behavior of dilute
$U$ atoms in the
cubic enviroments,\cite{impurex} thereby indicating the importance of lattice effects.
Second the nonlinear susceptibility as a function of pressure could be
measured;
we expect a
shoulder in $\chi_3 \sim {1\over T_0^3}$ that coincides with the 
observed development of Fermi liquid behavior in 
the resistance,\cite{100} the specific heat\cite{101},
and the magnetoresistance.\cite{103}

In conclusion we have performed a series of nonlinear susceptibility
measurements on the cubic heavy fermion
system $UBe_{13}$.  We find a small weakly
temperature-dependent anisotropy,  ${\Delta \chi_3 (\hat b)\over \chi_3}$,
in the nonlinear susceptibility
that is difficult to reconcile with the
quadrupolar Kondo scenario. These results  provide strong evidence for a
Kramers doublet ground-state in the $U^{3+}$ ions of $UBe_{13}$
and suggest a lattice mechanism for the observed non-Fermi liquid
behavior.  Further experiments have been proposed to test this
conjecture.

We thank D.L. Cox and A.M. Tsvelik for extensive discussions related to
this work.  P. Coleman is supported by NSF grant DMR-93-12138 and work
 at Los Alamos
was performed under the auspices of the UCDOE.

\gdef\journal#1, #2, #3, 1#4#5#6{       
    {\sl #1~}{\bf #2}, #3 (1#4#5#6)}

\def\pr{\journal Phys. Rev., }

\def\pra{\journal Phys. Rev. A, }

\def\prb{\journal Phys. Rev. B, }

\def\prc{\journal Phys. Rev. C, }

\def\prd{\journal Phys. Rev. D, }

\def\prl{\journal Phys. Rev. Lett., }

\def\jmp{\journal J. Math. Phys., }

\def\rmp{\journal Rev. Mod. Phys., }

\def\cmp{\journal Comm. Math. Phys., }

\def\adv{\journal Adv. Phys., }

\def\ap{\journal Adv. Phys., }

\def\np{\journal Nucl. Phys., }

\def\pl{\journal Phys. Lett., }

\def\apj{\journal Astrophys. Jour., }

\def\apjl{\journal Astrophys. Jour. Lett., }

\def\jpc{\journal J. Phys. C, }

\def\jetp{\journal Sov. Phys. JETP, }

\def\jetl{\journal Sov. Phys. JETP Letters, }

\def\phil{\journal Philos. Mag.}

\def\ssc{\journal Solid State Commun., }

\def\annp{\journal Ann. Phys. (N.Y.), }

\def\zpb{\journal Zeit. Phys. B., }

\def\jdp{\journal J.  Phys. (Paris), }

\def\jdc{\journal J.  Phys. (Paris) Colloq., }

\def\jpjap{\journal J. Phys. Soc. Japan, }

\def\physica{\journal Physica B, }

\def\phl{\journal Phil. Mag.,}

\def\phb{\journal Physica B, }

\def\jmmm{\journal J. Mag. Mat., }

\def\jmm{\journal  J. Mag. Mat., }

\def\sci{\journal Science, }

\newpage
$\ $
\begin{zigure}[here]
\epsfxsize=5.0truein
\hskip 1.0truein\epsffile{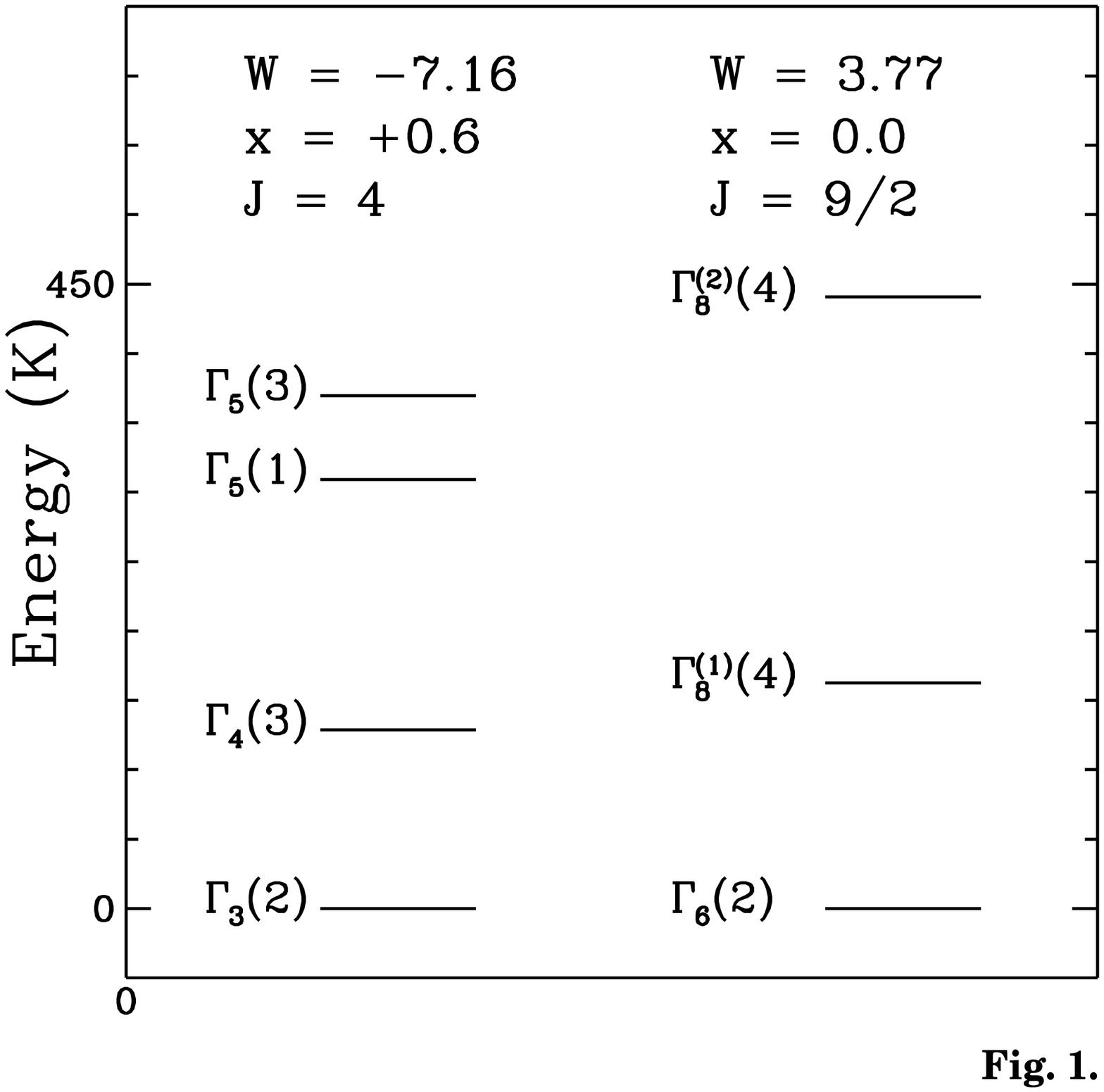}
\end{zigure}

\noindent
{\bf Fig. 1.}
The $J=4$ quadrupolar and the (b) $J={9\over 2}$
dipolar single-ion energy schemes for $UBe_{13}$
where 
the overall energy
scale and the level ordering are determined by a two-parameter fit to
the specific heat measurements of Felten et al.\cite{steglich}

\newpage
$\ $
\begin{zigure}[here]
\epsfxsize=3.0truein
\hskip 1.0truein\epsffile{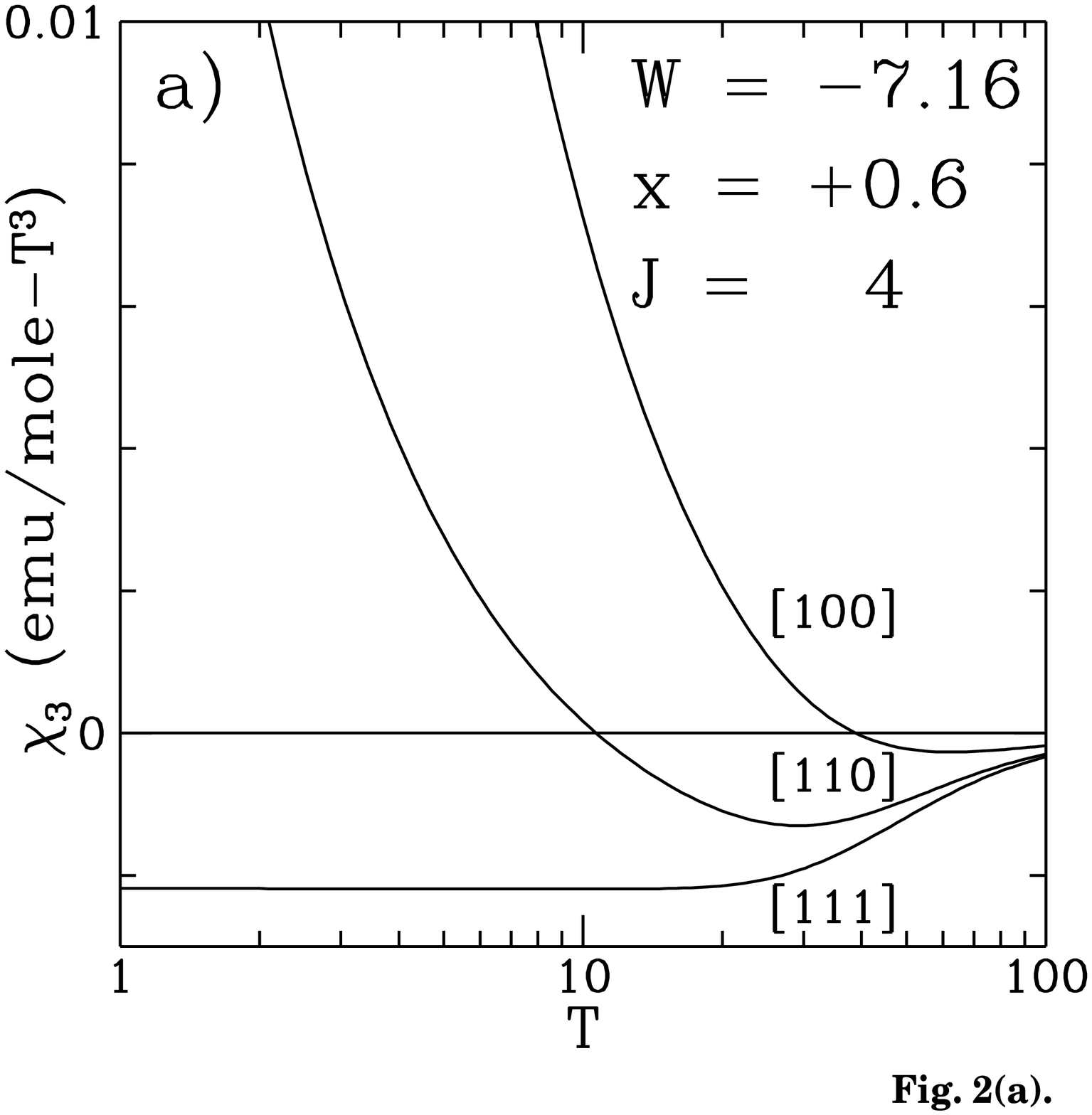}
\end{zigure}

$\ $
\begin{zigure}[here]
\epsfxsize=3.0truein
\hskip 1.0truein\epsffile{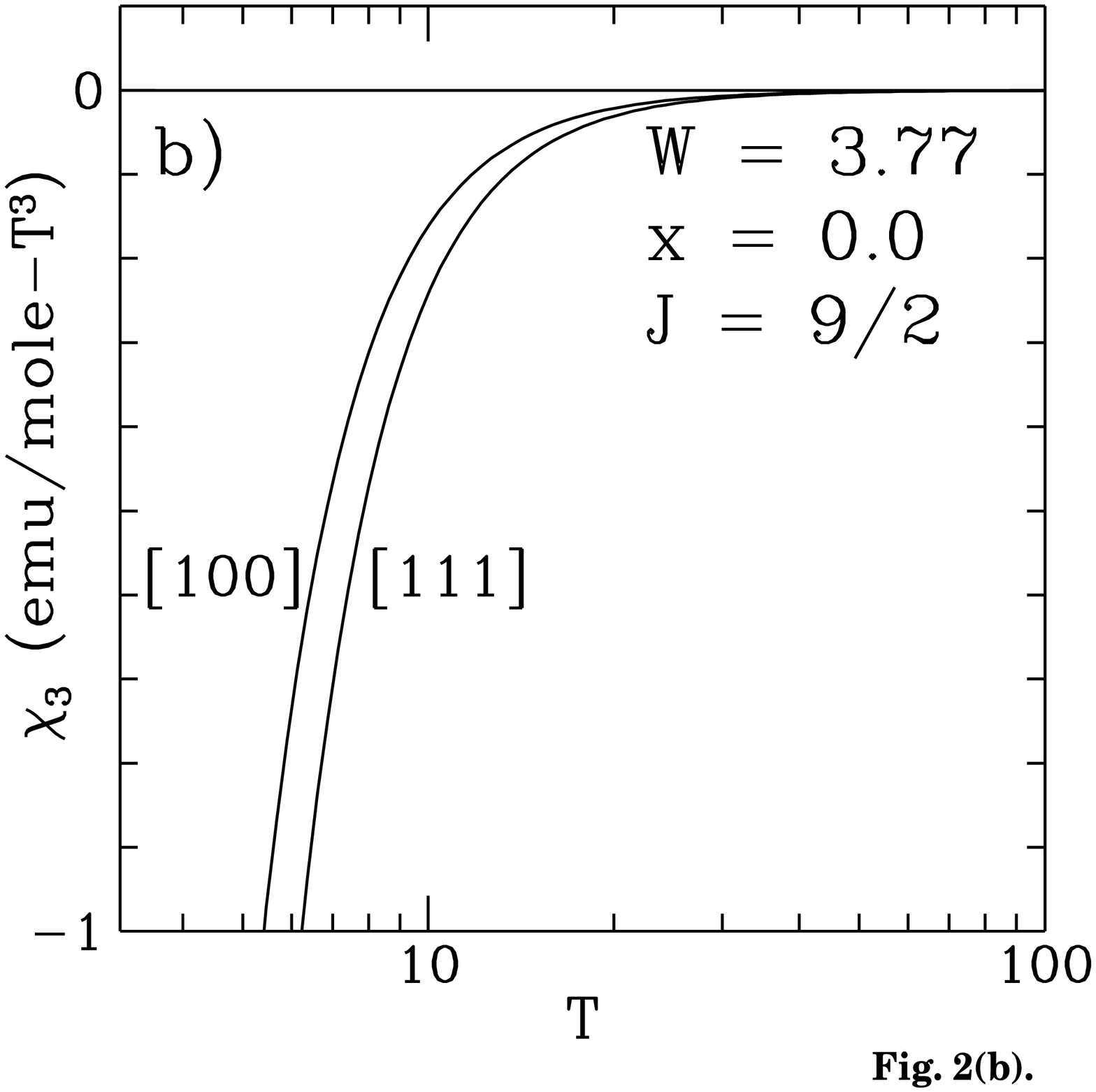}
\end{zigure}

\noindent
{\bf{Fig. 2.}
  The nonlinear susceptibility in the [100], [111]
and [110] directions for (a) the $J=4$ and  (b) the $J={9\over 2}$
energy schemes displayed in Figure 1.

\newpage
$\ $
\begin{zigure}[here]
\epsfxsize=5.0truein
\hskip 1.0truein\epsffile{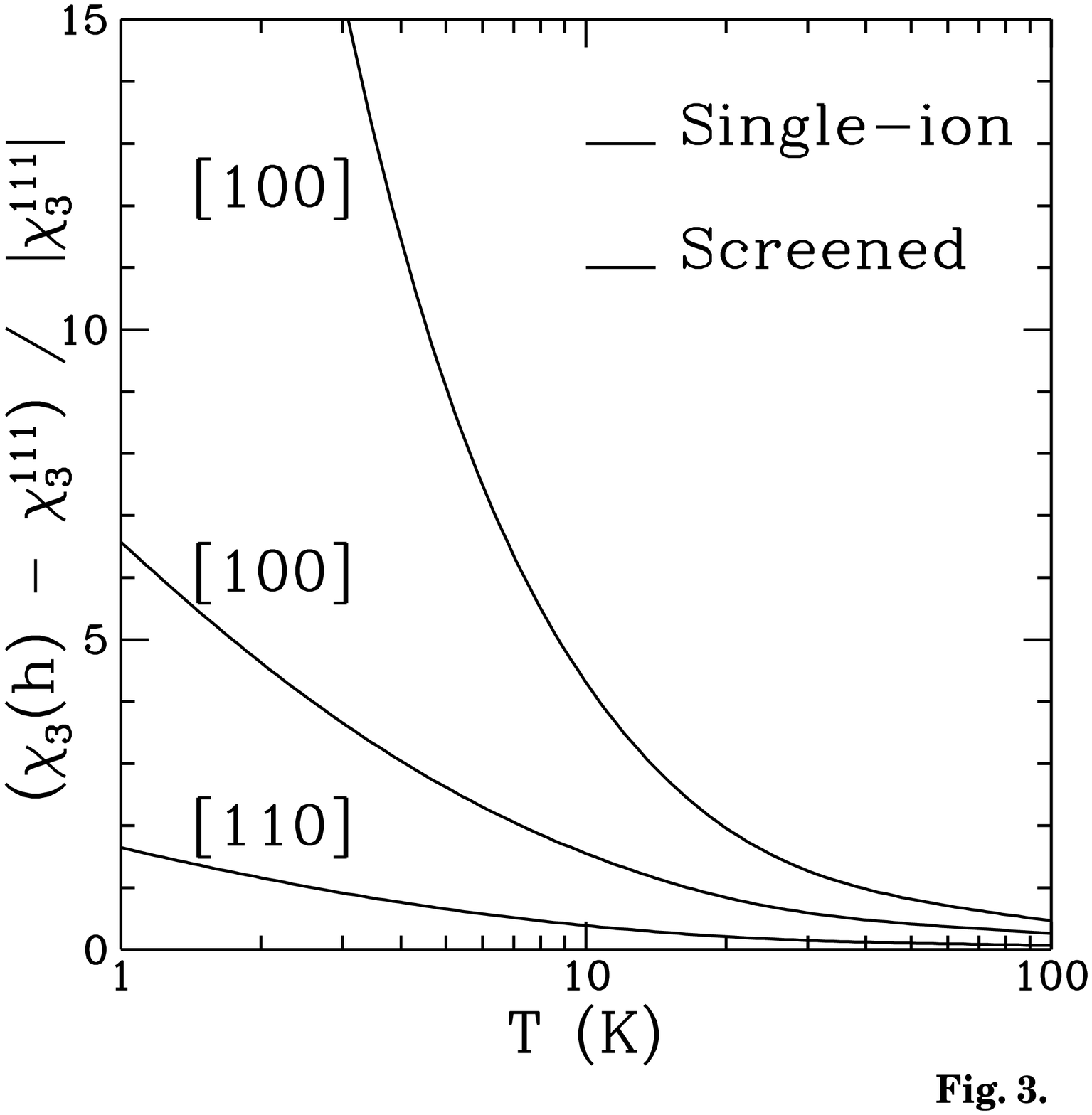}
\end{zigure}

\noindent
{\bf Fig. 3.} 
The anisotropic part of the nonlinear susceptibility
for the $J=4$ level scheme of Fig. 1.
The low-temperature ${\Delta \chi_3\over \chi_3}$ (dotted line) was determined by normalizing the 
single-ion anisotropy with the screening function
$f(T/T_0)$  from the solution of the two-channel Kondo problem;\cite{21} here the value
$T_0=1.5 K$ was extracted from the observed specific heat\cite{steglich}. 
For the $J={9\over 2}$ 
scheme of 
Figure 1 ${\Delta \chi_3(\hat b) \over \chi_3} =0$.

\newpage
$\ $
\begin{zigure}[here]
\epsfxsize=5.0truein
\hskip 1.0truein\epsffile{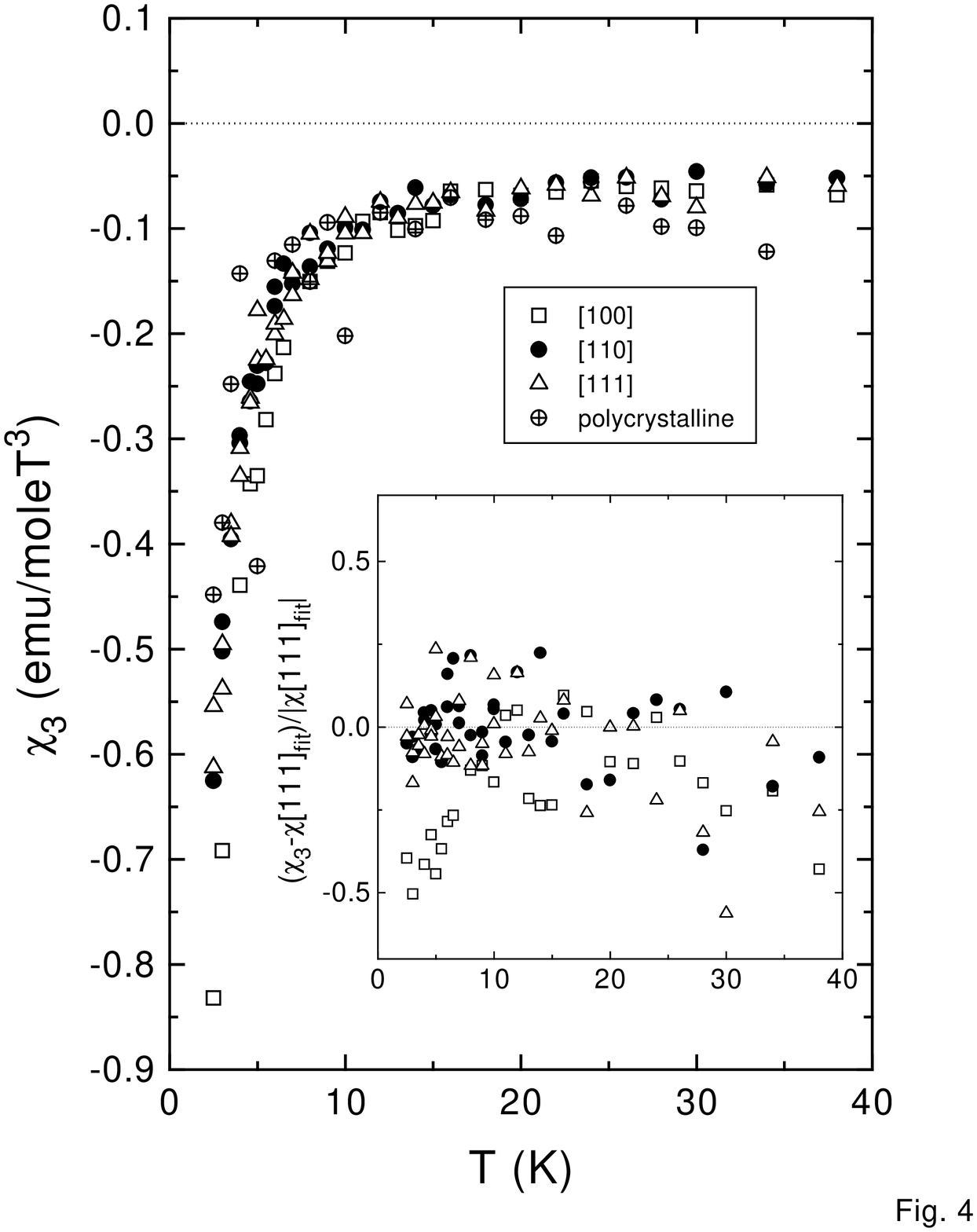}
\end{zigure}

\noindent {\bf Fig. 4.}
The measured nonlinear susceptibility ($\chi_3(\hat b)$)
and
$\Delta \chi_3 (\hat b) \over \chi_3$ (inset) for single-crystal and
polycrystalline
$UBe_{13}$.


\begin{references}
\bibitem{note}Typical valence fluctuations on energy scales of order
$100 - 1000$ meV would broaden the crystal-field levels to an
extent such that they would no longer be observable.
\end{references}
\end{document}